\documentclass[twocolumn,epjc3,keeplastbox]{svjour3}          
\RequirePackage[T1]{fontenc}
\smartqed                                   % flush right qed marks, e.g. at end of proof
\RequirePackage{graphicx}
\RequirePackage{mathptmx}      % use Times fonts if available on your TeX system
\RequirePackage{flushend}
\RequirePackage[numbers,sort&compress]{natbib}
\RequirePackage[colorlinks,citecolor=blue,urlcolor=blue,linkcolor=blue]{hyperref}
\usepackage{amsmath}
\usepackage{epsfig}
\usepackage{dcolumn}
\usepackage{bm}
\usepackage{slashed}
\usepackage{amsfonts}
\usepackage{amssymb}
\usepackage[english]{babel}
\usepackage{graphicx}
\usepackage{wrapfig}
\def\email#1{email: {\href{mailto:#1}{\nolinkurl{{#1}}}}}

\journalname{Eur. Phys. J. C}

\begin{document}

\title{Dynamical gravastars from the interaction between scalar fields and matter }

\author{A. Cabo Montes de Oca\thanksref{e1,addr1}
        \and
        D. Suarez Fontanella\thanksref{e2,addr1} 
        \and
        D. Valls-Gabaud\thanksref{e3,addr2}
}

\thankstext{e1}{\email{alejcabo@gmail.com} (corresponding author)}
\thankstext{e2}{\email{fontanella2405@gmail.com}}
\thankstext{e3}{\email{david.valls-gabaud@obspm.fr}}

\institute{Theoretical Physics Department, Instituto de Cibern\'{e}tica,
Matem\'{a}tica y F\'{\i}sica, Calle E, No. 309, Vedado, La Habana, Cuba. \label{addr1}
\and 
LERMA, CNRS UMR 8112, Observatoire de Paris,  61 Avenue de l'Observatoire, 75014 Paris, France. \label{addr2}
 }

\date{Received: date / Revised version: date}

\maketitle

\begin{abstract}
Gravastars are configurations of compact sin\-gu\-la\-ri\-ty-free gravitational objects which
are interesting alternatives to classical solutions in the strong gravitational field regime. 
Although there are no static star-like solutions of the Einstein-Klein-Gordon 
equations for real scalar fields, we show that \emph{dynamical gravastars} solutions 
arise through the direct interaction of a scalar field with matter. Two configurations 
presented here show that, within the internal zone, the scalar field plays
a role similar to a cosmological constant, while the scalar field decays 
at large distances as the Yukawa potential. 
Like classical gravastars, these solutions exhibit small values of the temporal metric component near a
transitional radial value, although this behaviour is not determined by the
de Sitter nature of the internal space-time, but rather by a slowly-varying scalar field.  
 The scalar field-matter interaction is able to define trapping forces that 
rigorously confine the polytropic gases to the interior of a sphere. At the
surface of these spheres, pressures generated by the field-matter interaction
play the role of "walls" preventing the matter from flowing out. These solutions
predict a stronger scattering of the accreting matter with respect to 
Schwarzschild black holes.
\end{abstract}
\authorrunning{Cabo Montes de Oca, Suarez Fontanella \& Valls-Gabaud}
\titlerunning{Dynamical gravastars}

\section{Introduction}
\label{s1}

The detection of gravitational waves by merging compact objects \cite{gw} and the imaging of
the environment around the supermassive compact object M87$^*$ 
\cite{photograph} has spawned renewed attention to alternative
solutions to the classic central singularities in collapsed matter
structures \cite{lrr} beyond General Relativity \cite{carballo}. However, current 
observations cannot, as yet, disentangle effects associated with the properties 
of the accretion flow from the ones produced by a strong gravitational field \cite{nonkerr}.

One of the most attractive alternatives are gravitational-vaccuum stars (gravastars), 
first proposed in references \cite{mazur,chapline,visser} as an extension of Bose-Einstein
condensates in gravitational systems, and which 
constitute a broad class of solutions which do not require exotic new physics 
as they are supported by negative pressure and present no singularities or
event horizons. Arbitrarily compact, they typically have
an internal region described by a de Sitter space-time 
which is matched at the horizon by an external Schwarzschild metric. The  matter is
concentrated at the boundary, which shows infinite surface tension, appear to
be dynamically stable \cite{chirenti}, and might match the observed lensed images
around M87$^*$ \cite{sakai}.

The present work identifies a new type of such solutions, which we call 
 \emph{dynamical gravastars} because the role of the de Sitter space in
repelling matter away from the origin of coordinates is played here by a scalar
field, a more dynamical object than the cosmological constant in the de Sitter space. 
The search for such solutions was first suggested by \cite{eloy,CE,bosonstar1,bosonstar2}, and 
in \cite{eloy,CE} it was argued that the Einstein-Klein-Gordon (EKG) equations can be solved exactly  in
the interior of a sphere with a scalar field configuration interacting with
gravity. There is a singularity at certain critical radius, and outside this
spherical zone, the solution is exactly the Schwarzschild space-time with a
zero value of the scalar field. These solutions, however, were found 
 in the sense of the Colombeau-Egorov generalized functions, which are still
in a clarification stage regarding its connections with real physical
solutions \cite{CETheory1,CETheory2}. If the solution
advanced in \cite{eloy,CE} is shown to have a physical meaning it could
furnish a further example against the validity of the Penrose Conjecture
\cite{penrose}. Within this framework, there are solutions 
 in which matter could be trapped in an interior
spherical region, outside of which the field reduces gradually (but within a
relatively short distance) to the classical Schwarzschild space-time with a
vanishing scalar field \cite{David}.

Further, in \cite{bosonstar0,bosonstar1,bosonstar2} it was argued that  the existence of a
static solution of the EKG equations became allowed thanks to the
assumed interaction of the scalar field with matter. Although the structure
resulted not to be closely resembling a neat gravastar, the discussion
suggested the possibility of obtaining further similar solutions.

Here we consider  physical systems constituted by a real
scalar field interacting with matter in two different forms:
an elastic solid and a polytropic gas. An important and already
mentioned element of the model is that the scalar field is considered 
interacting with the energy density, as it is relevant for reaching 
boson star solutions  \cite{bosonstar1,bosonstar2}. In  the two cases 
studied, this interaction is  implemented assuming that the source of the
scalar field is proportional to the particular matter energy density under
consideration. The Lagrangian also includes quadratic terms in the source,
which are incorporated ensuring a positive-definite energy density at any
point of the space-time.

First we write the  Einstein-Klein-Gordon equations for the
two forms of matter in terms of the scalar
field source $J(r)$, the energy density $\epsilon(r)$ and the pressure $P(r)$.
Second, the solutions must have configurations where the matter is 
rigorously confined within a spherical region, and we formulate 
 criteria for the existence of solutions with trapped matter configurations.

Numerical  solutions of the equations for each type of matter are found 
and all the field configurations show an internal region within which  all the matter  of
the system is exactly constrained to be inside a sphere of radius $r_{b}$,
at which point the pressure falls to zero.

The first type of dynamical gravastar solution  is associated with matter
defined by an elastic body with a constitutive relation between energy density
and pressure of the form $\epsilon(p)=\epsilon_{0}+\sigma p^{2}$, where the 
energy density grows with the square of the pressure.  The interaction between
the scalar field and matter is implemented  by an assumed proportionality of
the energy density $\epsilon(p(r))$ with the sources of the scalar field
$J(r)$. In this example, the equation coming from the Bianchi identity (which
implements the mechanical equilibrium condition for this system)  allows for
solutions showing a jump-like reduction of the pressure to zero values at some
radial point. Therefore,  this solution describes a
spherical elastic body with a boundary with the vacuum in the external
region. The scalar field decays in the faraway radial region like the
exponential Yukawa potential.

The second dynamical gravastar solution is to a polytropic gas with
a constitutive relation of the form  $p(r)= e^{-\gamma}  \epsilon(r)^{\gamma}$.
This case represents a gas of particles, and we expect that the
solutions might show a smooth decay of the pressure and density as the
radius increases. Surprisingly, we find that in addition to smooth solutions, there
are configurations in which the gas is rigorously
"trapped" within a spherical region by the interaction with the scalar field. 
 Again the equation associated to the Bianchi relations might develop
impulsive force densities acting like solid walls preventing particles of the
gas from flowing outside a trapped spherical region.

The two solutions presented here strongly
suggest the possibility of having regions in which the metric closely
approaches one with horizons, but not quite reaching the limit  to change its
signature for travellers passing though this region. That is, the space-time
does not unavoidably show self-trapped trajectories, attracting all the bodies
to a central singularity. In the external region, the solution is close
to the Schwarzschild space-time. These results argue, therefore, for the existence of
solutions of the EKG equations for interacting matter and scalar fields,
which truly constitute \emph{dynamical gravastars} alternatives to the classical solutions.

In section \ref{s2}, we present the general system of EKG equations including
direct interaction between matter and a scalar field, which are derived in \ref{appa}.  
Section \ref{s3} gives the 
general equations corresponding to matter  exactly confined within a spherical
region of radius $r_{b}$ , and their numerical solutions are given section \ref{s4} 
for elastic bodies. Section \ref{s5} presents solutions associated to a polytropic gas trapped by the
interaction with the scalar field inside a central spherical region. The
results are summarised in Section \ref{s6}. 

\section{The EKG equations including a field-matter interaction}
\label{s2}

The field equations to be solved are very close
in form to the ones considered in \cite{bosonstar1}, but we 
also derive them in  \ref{appa} for clarity. In their general form,
that is, without specifying the scalar field sources $J\,(r)$ and the
constitutive relation between the energy $\epsilon(r)$ and pressure $P(r)$,
the EKG equations are derived in  \ref{appa} as 
expressions (\ref{e1})-({\ref{e4}). Reproduced here, they are
% solution to ensure equations fit within one column
\begin{multline}
 E_{ekg}^{(1)}(r)    = {\frac{u_{,r}(r)}{r}}-{\frac{1-u(r)}{r^{2}}}\\+\frac
{1}{2}\left(u(r){\Phi_{,r}(r)}^{2}+\left(\Phi(r)+J(r)\right)^{2}\right)+\epsilon
(r)=0,\label{eecuaadim1} 
\end{multline}
\begin{multline}
E_{ekg}^{(2)}(r)  = \frac{u(r)}{v(r)}\frac{v_{,r}(r)}{r}-{\frac
{1-u(r)}{r^{2}}} \\ 
+\frac{1}{2}\left(-u(r){\Phi_{,r}(r)}^{2}+\left(\Phi(r)%
+J(r)\right)^{2}\right)-P(r)=0,\label{eecuaadim2}
\end{multline}
\begin{multline}
E_{ekg}^{(3)}(r)   = P_{,r}(r)+\left(\epsilon(r)+P(r)\right)\frac{v_{,r}(r)}%
{2v(r)}\\
-\left(J(r)+\Phi(r)\right)J_{,r}(r)=0,\label{eecuaadim3}
\end{multline}
\begin{multline}
E_{ekg}^{(4)}(r)    = J(r)+{\Phi}(r)-u(r)\text{ }{\Phi}^{\prime\prime}(r)  \\
-{\Phi}^{\prime}(r)\left(\frac{u(r)+1}{r} -\frac{r}{2}\left({\Phi
}(r)+J(r)\right)^{2}+\frac{\epsilon(r)-P(r)}{2}\right)=0, \label{eecuaadim4}
\end{multline}

%where in order to simplify the writing we have named the pressure $P^{\ast}$
%in the equations of \ref{appa}, as $P(r)$. 
The first two are the Einstein
equations associated to the temporal and radial diagonal components of the
mixed tensor Einstein equations. The third relation is the Bianchi identity
defined by the exact vanishing of the covariant divergence of the energy
momentum tensor. Physically, it represents the mechanical equilibrium of the
system. Finally, the fourth equation is the Klein-Gordon equation in the
curved space-time defined by the scalar and normal matter. We recall here 
the various notations employed below for the radial derivatives
of a function $f(r)$ :
\[
\frac{df(r)}{dr}=f_{r}(r)=f^{\prime}(r).
\]
}
We solve these equations adopting constitutive relations between the energy 
corresponding  to an
elastic body and to a polytropic gas. In the two situations the interaction
between matter and the sources of the
real scalar field is implemented  assuming a proportionality between
the scalar field sources and the energy density of the form 
\begin{equation}
J(r) \; = \; \alpha\text{ }\epsilon(r).
\end{equation}

\section{Gravastars and EKG equations including matter}
\label{s3}
In this section we search for general solutions of the  EKG
equations above (\ref{eecuaadim1}-\ref{eecuaadim4}) including the interaction between
 the scalar field and matter.
Specifically, we search for solutions resulting in a spherical 
gravastar-like configuration where the distribution of matter drastically ends at a radial distance
$r_{b}$. Outside this region, when $r>r_{b}$, the matter is absent.

Let us consider the scalar field sources and energy
density in equations (\ref{eecuaadim1})-(\ref{eecuaadim4}) to satisfy%
\begin{align}
J(r)  &  =\overline{J}(r)\text{ }\theta(r_{b}-r),\label{disc1}\\
\epsilon(r)  &  =\overline{\epsilon}(r)\text{ }\theta(r_{b}-r),\label{disc2}\\
P(r)  &  =\overline{P}(r)\text{ }\theta(r_{b}-r). \label{disc3}%
\end{align}
That is, the scalar field sources, the energy density and pressure are 
assumed to vanish for radii larger than $r_{b}$, outside the body.

Note that the proportionality between the scalar field source and
energy density functions conveys the interaction between the scalar field
and matter through
\[
\overline{J}(r)=\overline{\epsilon}(r),
\]
which is be valid for the two types of matter considered in the
following sections.

For the region $(0,$ $r_{b})$ the EKG equations in terms of the field
solutions in this neighborhood $u_{i}(r),v_{i}(r),\Phi_{i}(r)$ and $P(r)$,
take the form
\begin{multline}
\frac{1}{r}{\frac{du_{i}(r)}{dr}}-{\frac{1-u_{i}(r)}{r^{2}}}   = \\ -\frac{1}%
{2}\left(u_{i}(r)\left(\frac{d{\Phi}_{i}}{dr}{(r)}\right)^{2}+\left(\Phi_{i}(r)+\overline{J}%
(r)\right)^{2}\right)-\overline{\epsilon}(r),\label{int1}
\end{multline}
\begin{multline}
\frac{1}{r}\frac{u_{i}(r)}{v_{i}(r)}\frac{dv_{i}(r)}{dr}-{\frac{1-u_{i}%
(r)}{r^{2}}}   = \\ -\frac{1}{2}\left(-u_{i}(r)\left(\frac{d{\Phi}_{i}}{dr}{(r)}\right)^2
+\left(\Phi_{i}(r)+\overline{J}(r)\right)^{2}\right)+\overline{P}(r),\label{int2}
\end{multline}
\begin{multline}
0  =\frac{d\overline{P}(r)}{dr}+\left[\epsilon(r)+\overline{P}(r)\right]\frac
{1}{2v_{i}(r)}\frac{dv_{i}(r)}{dr}-\\ \left[\overline{J}(r)+\Phi_{i}(r)\right]\frac{d}%
{dr}\overline{J}(r),\label{int3}
\end{multline}
\begin{multline}
\overset{}{\text{ }\ \overline{J}}(r)+{\Phi}_{i}(r)-u_{i}(r)\text{ }{\Phi}%
_{i}^{\prime\prime}(r)   =  {\Phi}_{i}^{\prime}(r)\left[\frac{u_{i}(r)+1}%
{r} \right. \\ \left. -r\text{ }\left(\frac{{\Phi}_{i}(r)^{2}}{2}+\overline{J}(r){\Phi}_{i}%
(r)+\frac{\overline{J(}r)^{2}}{2}+\frac{\overline{\epsilon}(r)-\overline
{P}(r)\text{ }}{2}\right)\right]. \label{int4}
\end{multline}

%\begin{align}
%\frac{1}{r}{\frac{du_{i}(r)}{dr}}-{\frac{1-u_{i}(r)}{r^{2}}}  &  =-\frac{1}%
%{2}(u_{i}(r)(\frac{d{\Phi}_{i}}{dr}{(r))}^{2}+(\Phi_{i}(r)+\overline{J}%
%(r))^{2})-\overline{\epsilon}(r),\label{int1}\\
%\frac{1}{r}\frac{u_{i}(r)}{v_{i}(r)}\frac{dv_{i}(r)}{dr}-{\frac{1-u_{i}%
%(r)}{r^{2}}}  &  =-\frac{1}{2}(-u_{i}(r)(\frac{d{\Phi}_{i}}{dr}{(r))}%
%^{2}+(\Phi_{i}(r)+\overline{J}(r))^{2})+\overline{P}(r),\label{int2}\\
%0  &  =\frac{d\overline{P}(r)}{dr}+(\epsilon(r)+\overline{P}(r))\frac
%{1}{2v_{i}(r)}\frac{dv_{i}(r)}{dr}-(\overline{J}(r)+\Phi_{i}(r))\frac{d}%
%{dr}\overline{J}(r),\label{int3}\\
%\overset{}{\text{ }\ \overline{J}}(r)+{\Phi}_{i}(r)-u_{i}(r)\text{ }{\Phi}%
%_{i}^{\prime\prime}(r)  &  ={\Phi}_{i}^{\prime}(r)(\frac{u_{i}(r)+1}%
%{r}-r\text{ }(\frac{{\Phi}_{i}(r)^{2}}{2}+\overline{J}(r){\Phi}_{i}%
%(r)+\frac{\overline{J(}r)^{2}}{2}+\frac{\overline{\epsilon}(r)-\overline
%{P}(r)\text{ }}{2})). \label{int4}%
%\end{align}

Next, for radial values larger than $r_{b}$, that is, in the interval
$(r_{b},\infty)$ we consider the same set of equations (\ref{eecuaadim1}%
)-(\ref{eecuaadim4}), but where the scalar field sources, energy density
and pressure vanish exactly
\begin{multline}
\frac{1}{r}{\frac{du_{e}(r)}{dr}}-{\frac{1-u_{e}(r)}{r^{2}}}   = \\ 
-\frac{1}{2}\left(u_{e}(r)\left(\frac{d{\Phi}_{e}}{dr}(r)\right)^{2}+\Phi_{e}(r)^{2}\right),\label{ext1}
\end{multline}

\begin{multline}
\frac{1}{r}\frac{u_{e}(r)}{v_{e}(r)}\frac{dv_{e}(r)}{dr}-{\frac{1-u_{e}%
(r)}{r^{2}}}    = \\ -\frac{1}{2}\left(-u_{e}(r)\left(\frac{d{\Phi}_{e}}{dr}(r)\right)^{2}+\Phi_{e}(r)^{2}\right),\label{ext2}
\end{multline}
\begin{equation}
0    = 0\label{ext3}\\
\end{equation}
\begin{multline}
{\Phi}_{e}(r)-u_{e}(r)\text{ }{\Phi}_{e}^{\prime\prime}(r)    = \\ {\Phi}%
_{e}^{\prime}(r)\left(\frac{u_{e}(r)+1}{r}-r\frac{{\Phi}_{e}(r)^{2}}{2}\right) .
\label{ext4}%
\end{multline}

Here the fields $u_{e}(r), v_{e}(r), \Phi_{e}(r)$ indicate the solutions
of the above equations. Note that in this zone, the Bianchi identity is
automatically satisfied.

We now define an ansatz for the  solution of the EKG
equations we seek along the whole radial axis as
\begin{align}
u(r)  &  =u_{i}(r)\text{ }\theta(r_{b}-r)+u_{e}(r)\text{ }\theta
(r-r_{b}),\label{usol}\\
v(r)  &  =v_{i}(r)\text{ }\theta(r_{b}-r)+u_{e}(r)\text{ }\theta
(r-r_{b}),\label{vsol}\\
{\Phi}(r)  &  ={\Phi}_{i}\text{ }\theta(r_{b}-r)+{\Phi}_{e}(r)\text{ }%
\theta(r-r_{b}),\label{phisol}\\
{P}(r)  &  ={\overline{P}(r)}\text{ }\theta(r_{b}-r),\label{psol}\\
\epsilon(r)  &  =\overline{\epsilon}(r)\text{ }\theta(r_{b}-r),\label{esol}\\
J(r)  &  =\overline{J}(r)\text{ }\theta(r_{b}-r). \label{jsol}%
\end{align}

That is, the solution is assumed to coincide with the fields $u_{i}%
(r),v_{i}(r),\Phi_{i}(r)$ and $P(r)$ for points within the internal interval
\ $(0,$ $r_{b})$. In the external region $(r_{b},\infty)$ the solution is
chosen to be given by the external solution $u_{e}(r)$, 
$v_{e}(r)$, $\Phi_{e}(r)$.

\subsection{Criteria for solutions}

To find  gravastar-like solutions for the two forms of matter 
considered, we use the criterion employed in the theory of
generalised functions. In our case, it can stated as follows:
The system of differential equations $E_{ekg}^{(n)}(r)=0,$ for $n=1,2,3,4$
defined in equations (\ref{eecuaadim1})-(\ref{eecuaadim4}) are solved by the
fields $u(r),v(r),\Phi(r),P(r),\epsilon(r)$ and $J(r)$ specified in
(\ref{usol})-(\ref{jsol}), if all the integrals
\[
\int_{0}^{\infty}dr\ \Omega(r)\text{ }E_{ekg}^{(n)}(r)=0,\text{ }\ n=1,2,3,4,
\]
vanish for any arbitrarily-chosen test function $\Omega(r)$ pertaining to
$C^{\infty}.$

\subsection{The general solution for trapped matter region}

Let us first consider the equations $E_{ekg}^{(n)}(r)=0, \, n=1,2,4,$ \,that
is, excluding for the moment the Bianchi identity equation. In these three
equations there appear no derivatives of the suddenly-changing quantities 
$P(r),\epsilon(r)$ or $J(r)$  at the point $r_{b}$. Let us then 
assume that we determine the above-defined fields $u_{i}(r), v_{i}(r),
\Phi_{i}(r), P(r), \epsilon(r)$ and $J(r)$ that effectively solve the set of
internal equations (\ref{int1})-(\ref{int4}) in the entire interval $(0,r_{b})$,
in a way that the three fields $u_{i}(r)$, $v_{i}(r)$, 
$\Phi_{i}(r)$ have well-defined finite limits
\begin{align}
\lim_{r->rb}u_{i}(r_{b}) & =u_{i}(r_{b}), \\
\lim_{r->rb}v_{i}(r) & =v_{i} (r_{b}), \\
\lim_{r->rb}\Phi_{i}(r) & =\Phi_{i}(r_{b}),\\
\lim_{r->rb}\Phi^{\prime}_{i}(r) & =\Phi^{\prime}_{i}(r_{b}),
\end{align}
at the radial point $r_{b}.$

If the above conditions are met, let us also assume that the set of
equations for the external zone (\ref{ext1})-(\ref{ext4}) can also be  solved
in an external interval $\ (r_{b},\infty),$\ by fixing the ending values of
the interior solution at $r_{b}:$ $u_{i}(r_{b}),$ $v_{i}(r_{b}),\Phi_{i}%
(r_{b})$ and $\Phi_{i}^{\prime}(r_{b})$ as initial values for the equations
\ (\ref{ext1})-(\ref{ext4}) in the external region. These conditions at
$r_{b},$ then impose the continuity for the general solution of the quantities
$u(r)$ , $v(r)$, $\Phi(r)$ plus the continuity of the derivative of the scalar
field $\Phi^{\prime}(r),$ which can be fixed because the Klein-Gordon equation
is the only one of second order.

Consider now the antire radial axis as the union of small vicinities 
$(r_{b}-\delta,r_{b}+\delta)$ of the point $r_{b}$ and the two interior and
external intervals $(0,r_{b}-\delta)$ and $(r_{b}+\delta,\infty).$ Then, the
three equations can be written as
\begin{multline}
\int_{0}^{\infty}dr\, \Omega(r)\,E_{ekg}^{(n)}(r)   =\int_{0}^{r_{b}-\delta}dr\ \Omega(r)\,E_{ekg}^{(n)}(r) \\ 
+\int_{r_{b}-\delta}^{r_{b}+\delta}dr\, \Omega(r)\,E_{ekg}^{(n)}(r)+
\int_{r_{b}+\delta}^{\infty}dr\ \Omega(r)\text{ }E_{ekg}^{(n)}(r)\nonumber\\
  = \int_{r_{b}-\delta}^{r_{b}+\delta}dr\ \Omega(r)\text{ }E_{ekg}%
^{(n)}(r)=0,\; \; n=1,2,4,
\end{multline}
in which the integrals over the intervals $(0,r_{b}-\delta)$ and
$(r_{b}+\delta,\infty)$ identically vanish because the internal and external
fields satisfy the corresponding equations in such regions.

For the remaining integral is helpful to consider the definitions
of the ansatz for the three fields
\begin{align}
u(r)  &  =u_{i}(r)\text{ }\theta(r_{b}-r)+u_{e}(r)\text{ }\theta(r-r_{b}),\\
v(r)  &  =v_{i}(r)\text{ }\theta(r_{b}-r)+u_{e}(r)\text{ }\theta(r-r_{b}),\\
{\Phi}(r)  &  ={\Phi}_{i}\text{ }\theta(r_{b}-r)+{\Phi}_{e}(r)\text{ }%
\theta(r-r_{b}),
\end{align}
around the transition point $r_{b}$. It can be noted that only up to the first
derivatives of $u(r)$ , $v(r)$, and up to the second one of $\phi(r)$ appear
in the three equations $E_{ekg}^{(n)}(r)=0$ \, $n=1,2,4.$. However, the only
first derivatives involved of the continuous $u(r)$ and $v(r)$ can not
introduce any unbounded quantity in the interval $(r_{b}-\delta,r_{b}+\delta)$
due to the imposed continuity conditions. Further, the fact that  the
first derivative of the scalar field is continuous defines that the only
second derivative of the equations, which appears in the Klein-Gordon
equation, again is unable to introduce an unbounded term in the remaining integral.

Therefore, since $\delta$ is completely arbitrary, taking the limit
$\delta\rightarrow0$ shows  the three equations are satisfied%
\begin{equation}
\int_{0}^{\infty}dr\ \Omega(r)E_{ekg}^{(n)}(r)=0,\text{ \ \ \ \ \ \ \ \ \ \ }%
n=1,2,4.
\end{equation}

For the third equation, that is, the Bianchi identity, we decompose again the entire radial axis in the union of
small vicinity $(r_{b}-\delta,r_{b}+\delta)$ of the point $r_{b}$ and the two
interior and external intervals $(0,r_{b}-\delta)$ and $(r_{b}+\delta
,\infty).$ The condition for a solution of  the third equation takes the form
\begin{multline}
\int_{0}^{r_{b}-\delta}dr\ \Omega(r)E_{ekg}^{(3)}(r)+ \\ $ \; \; \;  \, $ \int_{r_{b}-\delta
}^{r_{b}+\delta}dr\ \Omega(r)\text{ }E_{ekg}^{(3)}(r)+ \int_{r_{b}+\delta
}^{\infty}dr\ \Omega(r)\text{ }E_{ekg}^{(3)}(r) = 0,\nonumber\\
\int_{r_{b}-\delta}^{r_{b}+\delta}dr\ \Omega(r)\text{ }E_{ekg}^{(3)}(r)   = 0,
\end{multline}
also because the interior and exterior fields, by construction, solve the four
equations in the internal and external zones. Therefore, the integration range is reduced
 to the interval of arbitrary width $2\,\delta.$

The remaining integral also can be rewritten as
\begin{align}
0 &  =\int_{r_{b}-\delta}^{r_{b}+\delta}dr\ \Omega(r)\text{ }\left(  \left(
\epsilon(r)+P(r)\right)  \frac{1}{2v(r)}\frac{dv(r)}{dr}\right)  +\nonumber\\
&  \int_{r_{b}-\delta}^{r_{b}+\delta}dr\ \Omega(r)\text{ }\bigg(  P^{\prime
}(r)-J^{\prime}(r)\left(J(r)+{\Phi}(r)\right)\bigg)  .\label{do}%
\end{align}

The first  integral vanishes because the interval width $2 \,\delta$ is
arbitrary and the integrand is a bounded function, since all the quantities
entering are finite. The remaining integral is considered for  the case in which the source \ $J(r)=-g$
$\epsilon(r)$  is not  constant as a function of the radius.

In this case  let us assume that the constitutive relations and the 
expression determining the interaction of the scalar field and
matter take the forms
\begin{align}
\epsilon\left(p(r)\right)  & = \; f(p(r)) \,,\\
J(r)  & = \; - g\text{ }\epsilon(r) \, .
\end{align}

Then, the integral (\ref{do})  can be  transformed to the form%

\begin{align}
0  & =\int_{r_{b}-\delta}^{r_{b}+\delta}dr\, \Omega(r)\,\bigg(
P^{\prime}(r)-J^{\prime}(r)(J(r)+{\Phi}(r))\bigg)  \nonumber\\
& =\int_{r_{b}-\delta}^{r_{b}+\delta}dr\, \Omega(r)\,P^{\prime}(r)\left(
1+g\big(-g\text{ }\epsilon(r)+{\Phi}(r)\big)\frac{\partial f(p)}{\partial p}\right).
\end{align}

In the integrand of the integral, it can be noted that  if the
pressure suddenly vanishes at $r=r_{b}$ the integral  does not vanish as
required, unless the function
\begin{equation}
Z(r)=1+g\bigg(-g\text{ }\epsilon(r)+{\Phi}(r)\bigg)\frac{\partial f(p)}{\partial p},
\end{equation}
also  reduces  to zero at  $r_{b}$.

For the third equation to be satisfied there is an additional
boundary condition for the ansatz to become a solution in the form
\begin{equation}
1+g\bigg(-g\text{ }\epsilon(r)+{\Phi}(r)\bigg)\frac{\partial f(p)}{\partial p} =0. \label{relation}
\end{equation}

We then arrive to the solution we were seeking: assuming that the interior and
exterior solutions can be obtained, that the fields $u,$ $v,$ ${\Phi}$ and
${\Phi}^{\prime}$ can be made continuous at the point $r_{b}$ and also that
the function $1+g (-g \epsilon(r)+{\Phi}(r))\frac{\partial}{\partial p}f(p)$ might also be 
fixed to vanish at the point $r_{b}$, the proposed
ansatz solution also satisfies the four EKG equations including matter
interacting with the scalar field.

In the next two  sections we find  numerical  solutions satisfying the
above  conditions for the two  types  of matter being trapped in the
central zone:  an elastic solid and a polytropic gas.

\section{The elastic body gravastar}
\label{s4}
The corresponding numerical solution of the equations
(\ref{eecuaadim1})-(\ref{eecuaadim4}) for  matter as
an elastic body assumes that the interaction between the scalar
field and matter isreflected in the adopted proportionality between the
energy density and scalar field sources $J(r)=g \epsilon(r).$

The energy density at the interior point is now defined as
\begin{align}
\epsilon(r)  & =\epsilon_{0}+\sigma\text{ }P(r)^{2}\\
& =f(P),
\end{align}
which reflects that the energy density of the body has a minimum at zero
pressure and grows quadratically with the pressure. Therefore, we are
interested in describing a spherical elastic body whose extension ends at the
 radius $r_{b}.$

For the evaluation of the searched solution satisfying the conditions
required in Section \ref{s3}, we simply follow the  procedure described  in the previous section.
After solving the equations in the interior region, the radial value $r_{b}$
is  determined finding the roots of the function $Z(r)$ which in 
this case is
\begin{align}
Z(r) &  =1+g\left(-g\text{ }\epsilon(r)+{\Phi}(r)\right)\frac{\partial}{\partial
P}f(P),\\
J(r) &  =\alpha\text{ }\epsilon(r)=\alpha(\epsilon_{0}+\sigma\text{ }%
P(r)^{2}).
\end{align}
Equations (\ref{int1})-(\ref{int3}) and (\ref{ext1})-(\ref{ext3})
are initially solved for $r<r_{b}$ $\ $ and $\ r>r_{b},$ respectively, and then 
an iterative procedure described below is  implemented to find
a solution showing a Yukawa-like behaviour for the scalar field in the distant regions.

The initial conditions for the equations are  chosen at a point 
very close to origin $r=\Delta$, and they are fixed  in the form
\begin{align}
\Phi(\Delta)  &  =1,\\
\Phi^{\prime}(\Delta)  &  =0,\\
u(\Delta)  &  =1,\\
v(\Delta)  &  =0.0085,\\
P(\Delta)  &  =0.2,\\
\Delta &  =0.000001.
\end{align}

The proportionality constants defining the interaction between scalar field
and matter and the elastic body constitutive relation were fixed to the values%
\begin{align*}
\alpha &  =3,\\
\epsilon_{0}  &  =0.931116,\\
\sigma &  =3.
\end{align*}

As it noted above, the solutions require a procedure for adjusting the
parameters. Assuming the  values of the parameters given above, the scalar
field solutions at large radial distances might appear by example, positive
(negative) at large radii. In this case increasing (decreasing) of the
$\epsilon_{0}$ parameter, it is always possible to  attain solutions for
which the scalar field tends to show negative (positive) values at large
radii. This property implies  an iterative procedure, in which
the scalar field of the arising solution can be made to decrease exponentially
at large radial values.

\begin{figure}[b]
\begin{minipage}{\columnwidth}
\centering
\includegraphics[width=\columnwidth]{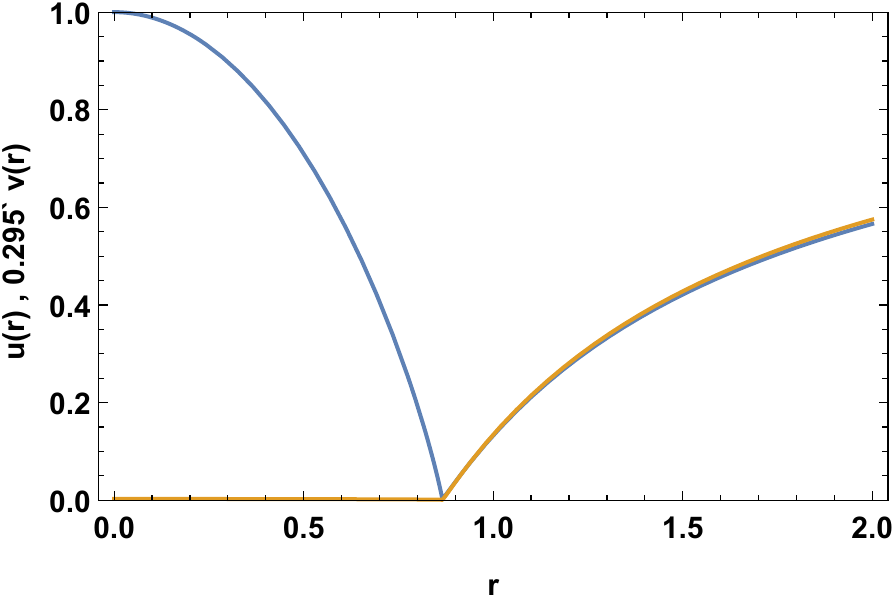}
\includegraphics[width=\columnwidth]{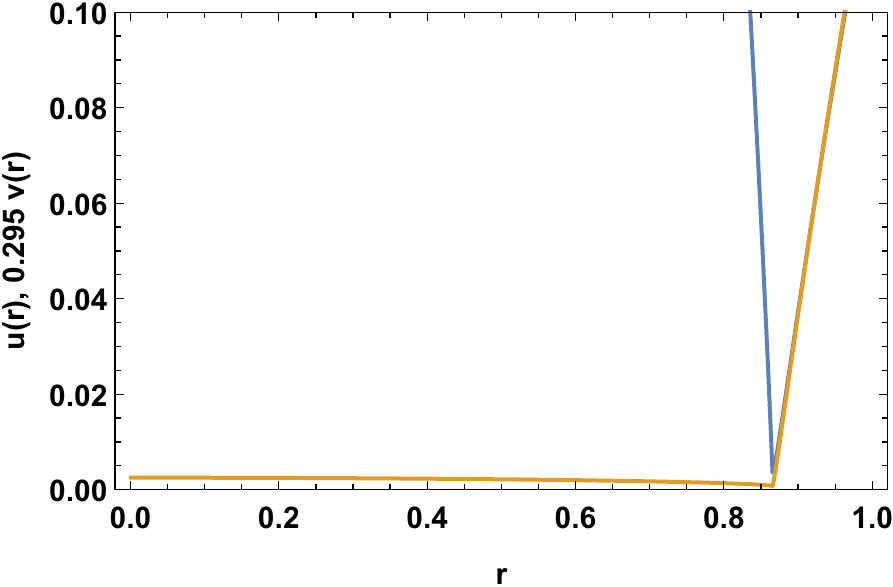}\
\end{minipage}
\caption{Radial dependence of $u(r)$ and $v(r)$
which define the metric. The top figure  shows $u(r)$ as the curve tending to
unity at the origin of coordinates and $v(r)$ as the one approaching a 
much smaller value at the origin. The bottom figure  shows the same curves
in detail at the boundary. Note that the potential $v(r)$ exhibits a maximum at the origin.
Therefore, in this case the gravitational potential decreases  away
form the center. This is a behaviour similar to the classical 
gravastars solutions  \cite{mazur,chapline,visser}.  }%
\label{EB-2}%
\end{figure}

Figure (\ref{EB-2} top) shows the dependence of
the radial diagonal component of the countervariant metric $u(r),$ in common
with the temporal component of the covariant metric. The metric
rapidly tends to be closely similar to the Schwarzschild one in the external zone, resembling the
structure of the classical solutions of this type
\cite{mazur,chapline,visser}. The solution shows a structure  very similar
 to a gravastar, as the minimum of $u(r)$ is 
very  close  to zero. In addition, the gravitational potential exhibits a
maximum at the centre of symmetry just as in the usual vacuum stars. 

\begin{figure}[t]
\begin{center}
\includegraphics[width=\columnwidth]{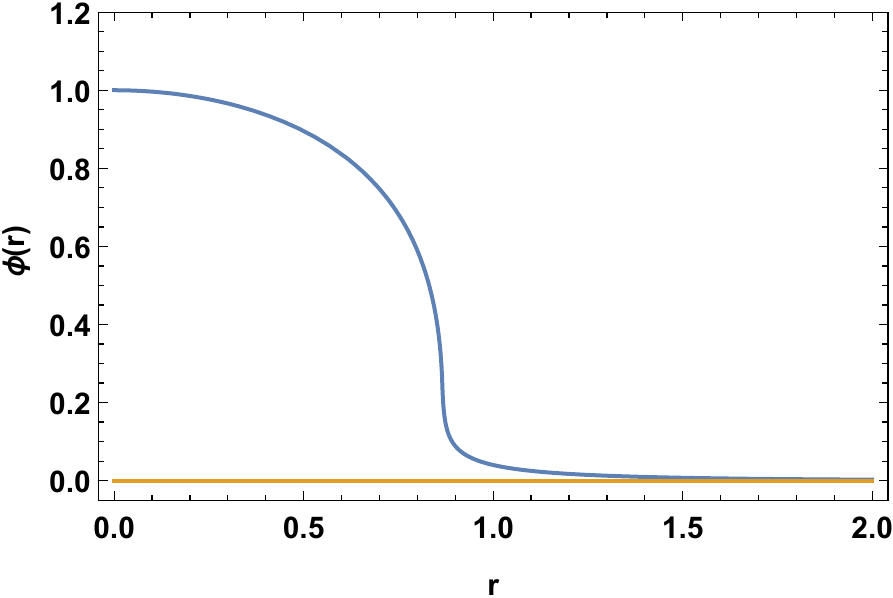}
\end{center}
\caption{The radial dependence of the scalar $\phi(r)$ for the elastic body
solution shows a
decay at infinity with an exponential Yukawa-like dependence. }%
\label{EB-4}%
\end{figure}

The temporal component of the covariant metric,  shown in detail
 in Figure (\ref{EB-2} bottom),  also approaches  zero at the surface of the
body. However, in this case, this field slightly increases to a
maximum at the origin. As before, in the external  regions the
radial behaviour of $u(r)$ and $v(r)$ tends to approach one to another (mainly
related by a multiplicative constant).  This describes the fact that the external 
zone asymptotically tends to be the Minkowski spacetime in the distant regions.

The scalar field behaviour is shown in Figure \ref{EB-4}, and tends to rapidly
decrease in the external region approaching the exponential
behaviour of the Yukawa potential.

\begin{figure}[t]
\begin{center}
\includegraphics[width=\columnwidth]{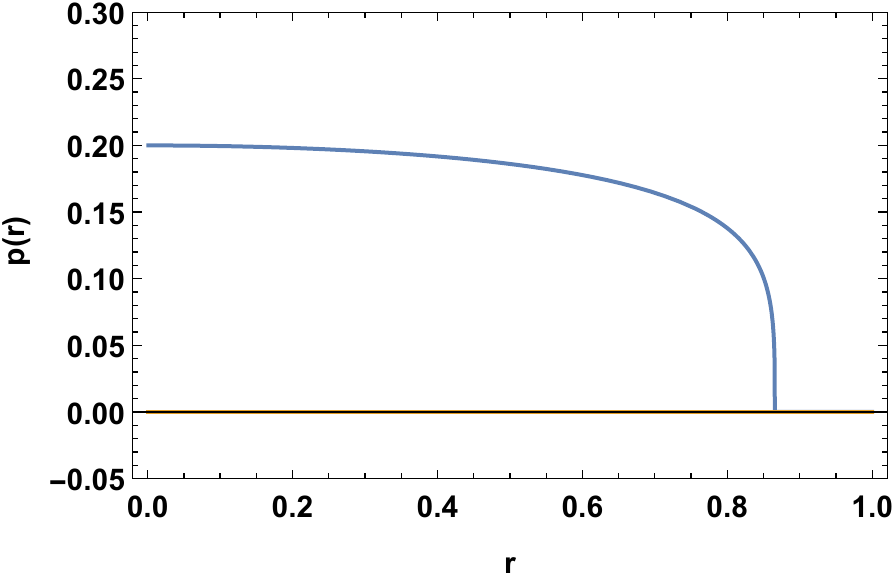}
\end{center}
\caption{The pressure of the configuration  tends to contract the
elastic body increasingly as the symmetry centre is approached. However, this
behaviour is curious, because the gravitational potential decreases when moving
away form the centre, attracting the matter to the boundary. This property
means that since the forces determined by the field-matter interaction oppose
the repulsing gravitational forces and overcome them, they lead to a net
contraction. Since the parameters can be varied, this property can be
different for other sets of  values. }%
\label{EB-5}%
\end{figure}

\begin{figure}[b]
\begin{center}
\includegraphics[width=\columnwidth]{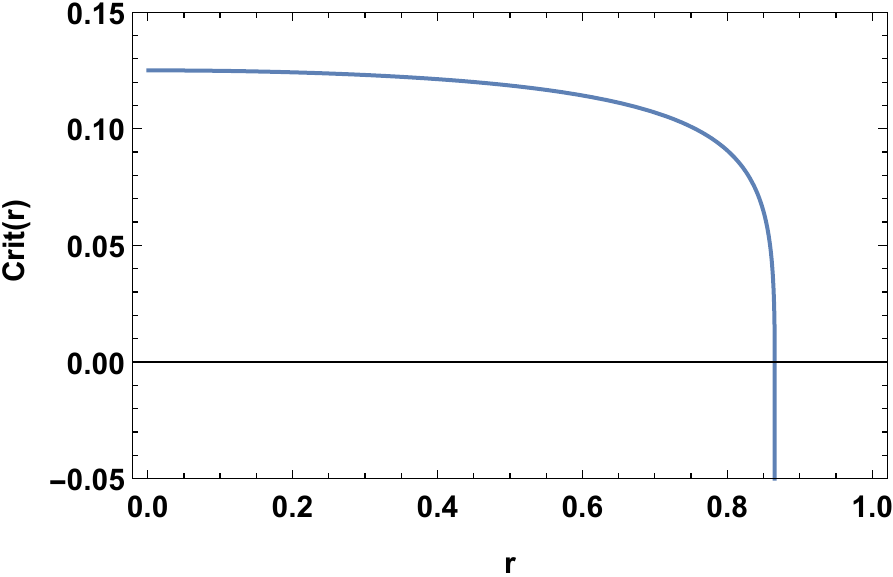}
\end{center}
\caption{The function $\mathrm{Crit(r)}=-Z(r) / (2\sigma g^2)$ for the case of the elastic
body which determines the value of $r_{b}$  at which it vanishes. }%
\label{EB-6}%
\end{figure}

The resulting radial dependence of the pressure in the elastic body is illustrated in
Figure \ref{EB-5}. As in the previous solution, the pressure grows
as the radial position tends to the centre of symmetry and drastically reduces to
zero at the boundary of the elastic body at the radial position $r_{b}.$
It should be noted that the equation of motion associated to the
Bianchi identify corresponds to a mechanical equilibrium of these 
systems. This equation is particularly relevant in defining the required
boundary conditions, and thus determining the concentrated pressures which act
over the boundary in this elastic body case. The internal pressure of
matter suddenly jumps at the radius $r_{b}$ to zero values at vacuum for
$r > r_{b}$ and Figure~\ref{EB-6} shows the zero crossing of the function $Z(r)$ 
which determines $r_{b}$. 
To conclude this section, it is of interest to note that the
 boundary pressures which appear here  can be expected to be present in the
just-discussed elastic body. This is
because parts of the solids are not expected to detach from the body at a free
boundary. However, in the next section we  discuss the case of a
polytropic gas, which is expected to occupy, through  diffusion
processes, all the volume accessible to it. In this case 
a force also appears, purely determined by the interaction of the scalar field with
matter, and which also confines the gas to the interior of a spherical region.

\section{The polytropic gas gravastar}
\label{s5}

The method for deriving this solution is similar to one
used in the previous  section. First, Equations (\ref{int1})-(\ref{int3}) 
and (\ref{ext1})-(\ref{ext3}) are solved by 
imposing continuity boundary conditions.

\begin{figure}[b]
\begin{center}
\includegraphics[width=\columnwidth]{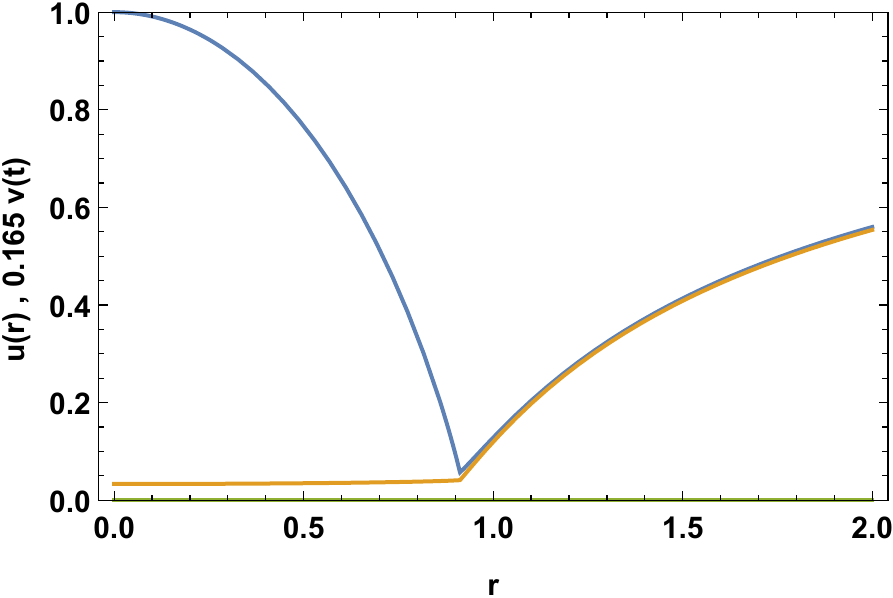}
\end{center}
\caption{The two metric functions. As before,
$u(r)$ is the curve approaching the value of unity at the origin. The gravitational
potential $v(r)$ in this case decreases towards the origin, and hence the
gravitational pressure contributes to increasingly compressing the gas when 
approaching  the origin. }%
\label{PG-2}%
\end{figure}

Similarly, the value of $r_{b}$ is determined finding the root with
respect to the radial coordinate of the function
\begin{align}
Z(r) &  =1+g\text{ }(-g\text{ }\epsilon(r)+{\Phi}(r))\frac{\partial}{\partial
P}f(P),\\
J(r) &  = -g \text{ }\epsilon(r),\label{coup}\\
P(r) &  =\exp(-\gamma)\text{ }\epsilon(r)^{\gamma},\label{const}\\
f(P) &  =\exp(1)\text{ }P^{\frac{1}{\gamma}}\\
\gamma &  =7.13571\\
g &  =0.897102,\label{a}%
\end{align}
where (\ref{const}) is the constitutive relation between the energy and
pressure of a polytropic gas and (\ref{a}) is the constant defining the matter
and scalar field interaction defined by relation (\ref{coup}).

The iterative process of varying the parameters (in this case the
absolute value of the constant $\alpha$) is performed step by step to
reach  a solution which shows a scalar field decaying as the Yukawa potential
at large radii.

\begin{figure}[t]
\begin{center}
\includegraphics[width=\columnwidth]{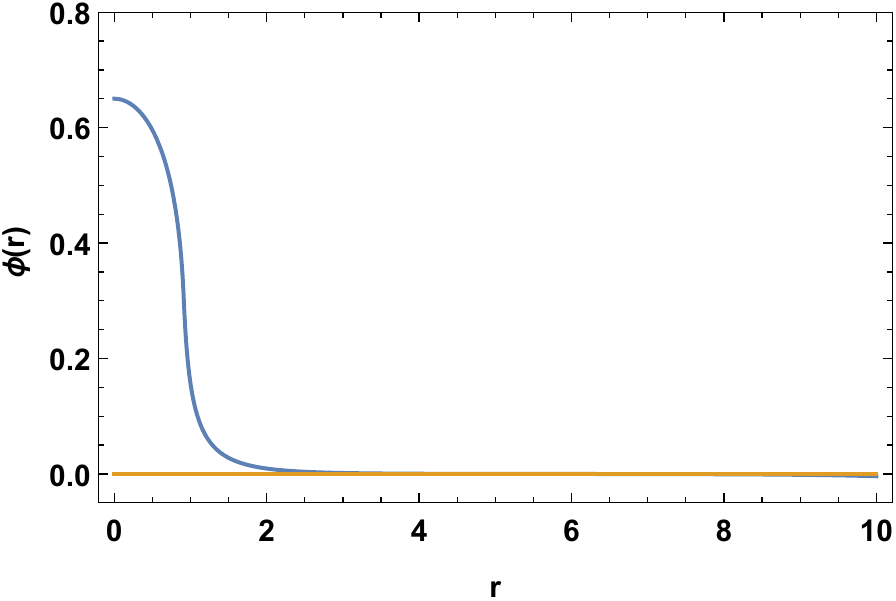}
\end{center}
\caption{The scalar field in this case also shows the bell-like behaviour as
a function of  radius, which again tends to the Yukawa potential form at
large distances. }%
\label{PG-3}%
\end{figure}

As  remarked at the end of the previous section, an interesting question
in connection with the polytropic gas case is the following. Up to what extent a solution
exists showing the gas completely trapped within a spherical region with the
boundary surface at $r=r_{b}$?  The elastic solid
situation can be expected to show a vanishing pressure at the outside because
the body is a   solid-like one, but the polytropic gas could perhaps not 
have the gas fully confined to a spherical region.

 \begin{figure}[b]
\begin{center}
\includegraphics[width=\columnwidth]{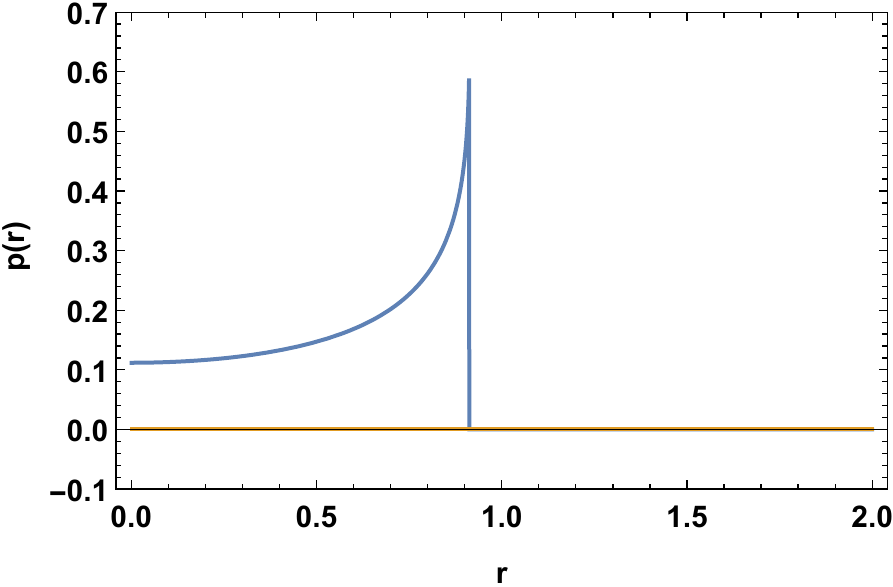}
\end{center}
\caption{Radial dependence of the pressure  for the
polytropic gas system. For this type of matter,  and the chosen
parameters, the pressure increases as the distance to the boundary becomes
smaller.
This is at variance with the case of the  elastic body, as shown in Fig. \ref{EB-5}.}%
\label{PG-4}%
\end{figure}

In fact the polytropic gas also shows solutions in which the gas is rigorously
confined within a spherical region of radius $r_{b}$.

As discussed in the case of the elastic body, the initial conditions were fixed at a
point $r=\Delta$ such that
\begin{align}
\Phi(\Delta)  &  =0.65,\\
\Phi^{\prime}(\Delta)  &  =0,\\
u(\Delta)  &  =1,\\
v(\Delta)  &  =0.2,\\
\epsilon(\Delta)  &  =2,\\
\Delta &  =0.000001.
\end{align}

\begin{figure}[t]
\begin{center}
\includegraphics[width=\columnwidth]{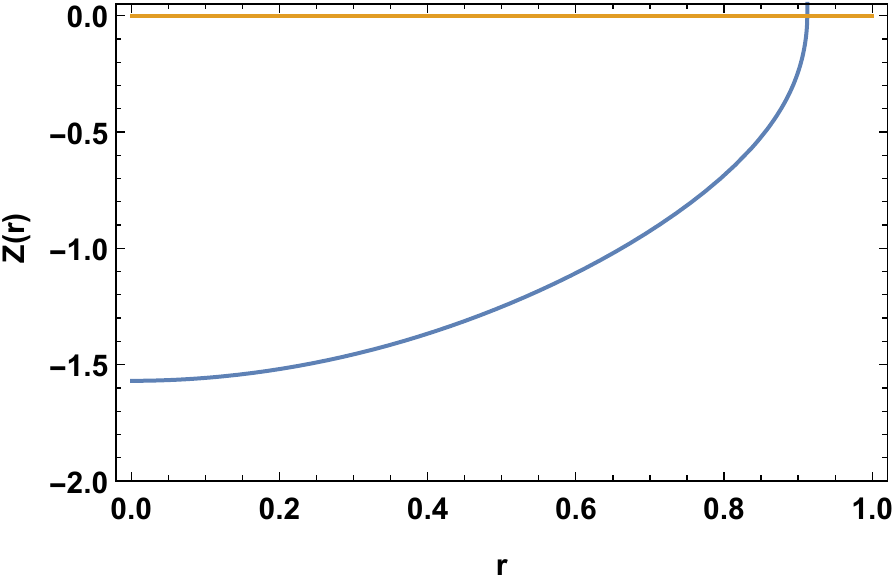}
\end{center}
\caption{The  $Z(r)$ function (after being multiplied by a constant) for the polytropic
gas system, which defines  $r_{b}$ as the radius when it reaches the zero value. }%
\label{PG-5}%
\end{figure}

The quantities $u(r)$, $v(r)$, $\Phi$, $P(r)$, $\epsilon(r)$ and $Z(r)$  are
shown in Figures \ref{PG-2}-\ref{PG-5}. These solutions for a 
the polytropic gas show a behaviour similar to the previously-discussed ones 
of elastic bodies.  In both of them a sudden change at the boundary of the
pressure to a zero value at vacuum is  present and the space-time tends to
be the Schwarzschild one at large distances, after a rapid transition in which
the scalar field decreased. The asymptotic behaviour  of the scalar field in the two cases
is a Yukawa-like exponential .

It is an interesting and unexpected outcome that the case of a polytropic gas also presents
wall-like forces which rigorously confine the gas within the interior of a
spherical region. This result is a direct consequence of the assumed 
interaction between the scalar field with matter: the concentrated
impulsive forces determined by the interaction terms in the fourth divergence
of the energy momentum tensor are the confining forces acting like a spherical
``wall" over the gas of particles.

This  matter-scalar field interaction is implemented by the sources of the
scalar field which are proportional to the matter energy density. That is,  the
confining forces over the polytropic gas matter are generated by the
interaction of the scalar field with matter.

\section{Summary}
\label{s6}
We have found new examples of gravastar-like field configurations in General Relativity which
are solutions of the EKG equations including matter. 
Their existence is allowed by the presence of a direct interaction between the
scalar field and matter. Within the internal zone, the scalar field plays
a role similar to a cosmological constant, and the solutions can be considered as
dynamically-generated gravastars. The metrics have a region in which its
temporal component take values close to zero near a radial distance $r_{b}$.
For larger  radial values the metric tensor rapidly tends to the
Schwarzschild case. The configurations considered show a scalar field behaving
as the Yukawa potential at large radii and the scalar
field-matter interaction is able to define trapping forces that 
rigorously confine the polytropic gases to the interior of a sphere. In the
surface of these spheres, pressures generated by the field-matter interaction
play the role of "walls" preventing the matter from flowing out. 

Finally, it is useful to remark that the relevance of the
interaction of matter with the scalar field suggests the possible important
role of string theory for the constituency of such structures. This idea comes
from the fact that the Yukawa interaction of the scalar moduli fields with
fermion matter fields makes completely natural the presence of such
interactions in  field theory approximations of  string theory \cite{lust,alexis}.

It will be interesting to investigate  in detail image formation under  these configurations  
even under simplified accretion disc physics, such as the analysis carried out in \cite{nonkerr}. Due to the 
structure of these \emph{dynamical gravastars}, we expect 
that they will show evidence for a stronger scattering of the accreting matter 
than the one found in  Schwarzschild black holes. This possibility
comes from the presence of matter directly at the entrance of the captured
beams in the interior regions \cite{olivares}.

\section*{Acknowledgments}

The authors acknowledge support received from the Office of External
Activities of ICTP (OEA), through the Network on \textit{Quantum Mechanics,
Particles and Fields} (Net-09). and from the Service de  coop\'eration et
d'action culturelle of the Embassy of France in Cuba. 
They are also indebted to Eric Gourgoulhon (LUTh, Paris Observatory, France),  
Hector Olivares (Radboud University, Nijmegen, The Netherlands) 
and Carlos Garc\'{\i}a  (Faculty of Physics, Havana University,  Cuba) for
fruitful discussions. 

\section*{Data Availability Statement} 
This manuscript has no associated data or the data will not be deposited. [Authors’ comment: The results in the paper are based solely on the authors’ computations, and there is no further associated data. In case of interest please contact the corresponding author.]

\section*{Open Access} This article is licensed under a Creative Commons Attribution 4.0 International License, which permits use, sharing, adaptation, distribution and reproduction in any medium or format, as long as you give appropriate credit to the original author(s) and the source, provide a link to the Creative Commons licence, and indicate if changes were made. The images or other third party material in this article are included in the article’s Creative Commons licence, unless indicated otherwise in a credit line to the material. If material is not included in the article's Creative Commons licence and your intended use is not permitted by statutory regulation or exceeds the permitted use, you will need to obtain permission directly from the copyright holder. To view a copy of this licence, visit 
\url{http://creativecommons.org/licenses/by/4.0/}.
Funded by SCOAP$^{3}$.

\appendix

\section{The general field equations}
\renewcommand{\thesection}{\Alph{section}}
\label{appa}

In this Appendix we reproduce for bookkeeping reasons the derivation of the
Einstein-Klein-Gordon equations that were presented in \cite{bosonstar1}.

The physical system includes matter defined by a scalar field and usual matter showing
a  constitutive relation expressing the energy density as a general
function of the pressure.

The metric is associated to the squared proper time interval 
\begin{align}
ds^{2}  &  =\mathit{v}(\rho){dx^{0}}^{2}-u(\rho)^{-1}d\rho^{2}-\rho
^{2}(sin^{2}\theta d\varphi^{2}+d\theta^{2}) \, ,\\
x^{0}  &  =ct\, , \; \; \; x^{1}=\rho\, , \; \; \; x^{2}   = \varphi \, ,\; \; \; x^{3}\equiv\theta \, .
\end{align}
The Einstein tensor $G_{\mu\nu}$ in terms of functions $u,v$ and the radial
variable $\rho$ can be evaluated as 
\begin{align}
{G_{0}^{0}}  &  ={\frac{u^{\prime}}{\rho}}-{\frac{1-u}{\rho^{2}}},\\
G_{1}^{1}  &  ={\frac{u}{v}}{\frac{v^{\prime}}{\rho}}-{\frac{{1-u}}{\rho^{2}}%
} \, ,\\
{G_{2}^{2}}={G_{3}^{3}}  &  ={\frac{u}{2v}}{v^{\prime\prime}}+{\frac
{uv^{\prime}}{4v}(\frac{u^{\prime}}{u}-\frac{v^{\prime}}{v})}\nonumber\\
&  +{\frac{u}{2\rho}(\frac{u^{\prime}}{u}+\frac{v^{\prime}}{v})} \, .
\end{align}
In what follows we use various notations for a derivative of a function
$f(x)$ as
\[
\frac{df(x)}{dx}=f_{,x}(x) \, = \, f^{\prime}(x) \, .
\]

The physical system interacting with gravity is considered to be made of 
a scalar field and a  material body, both with spherical symmetry. The scalar field is also assumed to 
 interact linearly with an external source associated with it. Further, the
source field is considered to be proportional to the body energy density.
This assumption  introduces the interaction of the scalar field with
matter. It should be noted that in most of the former studies of the EKG equations including
matter the substance had not been considered as directly interacting with the
field \cite{jet,lidd1,lidd2,lidd3,urena1,urena2}.

The action of the field takes the form%
\begin{equation}
{S}_{mat-\phi} \, = \, \int L\sqrt{-g}d^{4}x \, ,
\end{equation}
with a Lagrangian density given by
\begin{equation}
{L}\, = \, {\frac{1}{2}}\left(g^{\alpha\beta}{\phi}_{,\alpha}{\phi}_{,\beta}+m^{2}{\phi
}^{2}+2\text{ }j(\rho)\text{ }\phi+j^{2}(\rho)\right) \, . \label{denslag}%
\end{equation}

This Lagrangian determines an energy momentum of the form
\begin{equation}
(T_{mat-\phi})_{\mu}^{\nu} \, = \, -\frac{\delta_{\mu}^{\nu}}{2}\left(g^{\alpha\beta}{\phi}_{,\alpha}{\phi}_{,\beta}+m^{2}{\phi}^{2}+2\text{ }j(\rho)\text{ }\phi+j^{2}(\rho)\right) \, ,\nonumber
\end{equation}
which can be added to the energy momentum tensor of the matter
(\cite{synge}):
\begin{equation}
(T_{e,p})_{\mu}^{\nu}\, = \, p\,\delta_{\mu}^{\nu}-u^{\nu}u_{\mu}(p+\epsilon) \, ,
\end{equation}
so as to write the total energy momentum tensor as
\begin{align}
T_{\mu}^{\nu}  &  =-\frac{\delta_{\mu}^{\nu}}{2}\left(g^{\alpha\beta}{\phi
}_{,\alpha}{\phi}_{,\beta}+m^{2}{\phi}^{2}+2j(\rho)\text{ }\phi +j^{2}(\rho)\right)\nonumber\\
&  \text{ \ \ \ }+g^{\alpha\nu}{\phi}_{,\alpha}{\phi}_{,\mu}+p \, \delta_{\mu}^{\nu}-u^{\nu}u_{\mu}(p+e) \, .
\end{align}

Since static field configurations are being searched for, the four-velocity
reduces to the form in the local rest system at any spatial point
\begin{equation}
u^{\mu}(t,r)=(1,0,0,0) \, .
\end{equation}

After these definitions, the Einstein equations for the considered system can
be written as usual
\begin{equation}
G_{\mu}^{\nu}=G\hspace{0.1mm}\,\, T_{\mu}^{\nu} \, , \label{Einstein}
\end{equation}
in which both tensors are diagonal and the gravitational
constant has the value
\begin{equation}
G \, = \, 8\pi \; l_{P}^{2} \, ,
\end{equation}
in terms of the Planck length $l_{P}=1.61\times10^{-36}$ m.

The non-vanishing Einstein equations determined by the diagonal terms take the form
\begin{multline}
{\frac{u^{\prime}}{\rho}}-{\frac{1-u}{\rho^{2}}}   = \\ 
\hfill -G\left[\frac{1}{2}\left(u\phi_{,\rho}^{2}+m^{2}\phi^{2}+2 j \phi+j^{2}\right)+\epsilon\right] \, ,\\
{\frac{u}{v}}{\frac{v^{\prime}}{\rho}}-{\frac{{1-u}}{\rho^{2}}}  = \\
\hfill G\left[\frac{1}{2}\left(u\phi_{,\rho}^{2}-m^{2}\phi^{2}-2 j \phi -j^{2}\right)+p\,\right] \, ,\\
{\frac{\rho^{2}u}{2}}{v^{\prime\prime}}+\frac{\rho^{2}}{4}u\text{ }v^{\prime
}({\frac{{u}^{\prime}}{u}-}\frac{{v}^{\prime}}{v})+\frac{\rho}{2}u\text{
}v^{\prime}   = \\ 
\hfill G\left[\frac{1}{2}\left(u\phi_{,\rho}^{2}+m^{2}\phi^{2}+2j \phi+j^{2}\right)+p\,\right] \, ,\\
{\frac{\rho^{2}u}{2}}{v^{\prime\prime}}+\frac{\rho^{2}}{4}u v^{\prime}({\frac{{u}^{\prime}}{u}-}\frac{{v}^{\prime}}{v})+\frac{\rho}{2}u v^{\prime}    = \\ 
G\left[\frac{1}{2}\left(u\phi_{,\rho}^{2}+m^{2}\phi^{2}+2 j \phi+j^{2}\right)+p\,\right] \, .
\end{multline}

These are four equations, the last two of which are identical. Thus, there are
three independent Einstein equations in the problem. However, the third and the
equivalent fourth one can be substituted by a simpler relation which comes from
the Bianchi identities (\cite{synge}):
\begin{equation}
G_{\mu\text{ };\text{ }\nu}^{\nu}=0, \label{Bianchi}%
\end{equation}
where the semicolon indicates the covariant derivative of the tensor
$G_{\mu\text{ }}^{\nu}.$ After assuming  the Einstein
equations (\ref{Einstein}) are satisfied, the $G_{\mu}^{\nu}$ tensor in (\ref{Bianchi}) can
be substituted by the energy momentum tensor $T_{\mu}^{\nu}$ leading to the
relation
\begin{equation}
-(\phi+j)\text{ }j^{\prime}+p^{\prime}+\frac{v^{\prime}}{2v}(p+e)=0,
\end{equation}
which are dynamical equations for the energy, the pressure and the scalar
field, substituting the two equivalent Einstein equations being associated to
the two angular directions, related to the mechanical equilibrium of the system.

The last of the equations of movement for the system is the Klein-Gordon one
for the scalar field. It can be obtained by imposing the vanishing of the
functional derivative of the action $S_{mat-\phi}$ with respect to the field
\begin{align}
\frac{\delta S_{mat-\phi}}{\delta\phi(x)}  &  \equiv\frac{\partial}{\partial
x^{\mu}}\frac{\partial L}{\partial\phi_{,\mu}}-\frac{\partial L}{\partial\phi
}\nonumber\\
&  \equiv\frac{1}{\sqrt{-g}}\frac{\partial}{\partial x^{\mu}}(\sqrt{-g}%
g^{\mu\nu}\phi_{,\nu})-m^{2}\phi-j\nonumber\\
&  =0,
\end{align}
a relation that after using the temporal and radial Einstein equations in
(\ref{Einstein}) can be written in the form
\begin{multline}
j(\rho)+m^{2}\phi(\rho)-u(\rho)\text{ }\phi^{\prime\prime}(\rho) \, = \\
\phi^{\prime}(\rho)\left(\frac{u(\rho)+1}{\rho}-\rho\text{ }G\text{ }\left(\frac{m^{2}\phi(\rho)^{2}}{2}+j(\rho)\text{ }\phi(\rho)+ \right. \right. \\
\left. \left. \frac{j(\rho)^{2}}{2}+\frac{e(\rho)-p(\rho)}{2}\right)\right). \label{escalar}%
\end{multline}

Therefore, the relevant EKG equations for the problem can be reduced to
\begin{multline}
{\frac{u^{\prime}(\rho)}{\rho}}-{\frac{1-u(\rho)}{\rho^{2}}}   \, = \\ 
-\,G\left[\frac{1}{2}\left(u\phi_{,\rho}^{2}(\rho)+m^{2}\phi(\rho)^{2}+2j(\rho)\phi(\rho)+j(\rho)^{2}\right)+e(\rho)\right] \, ,\\
{\frac{u(\rho)}{v(\rho)}}{\frac{v^{\prime}(\rho)}{\rho}}-{\frac{{1-u}(\rho)}{\rho^{2}}}    \, = \\ 
G\left[\frac{1}{2}\left(u\phi_{,\rho}^{2}(\rho)-m^{2}%
\phi(\rho)^{2}-2\text{ }j(\rho)\text{ }\phi(\rho)-j(\rho)^{2}\right)+p(\rho)\,\right], ,\\
j(\rho)+m^{2}\phi(\rho)-u(\rho)\text{ }\phi^{\prime\prime(\rho)}  \\
=\phi^{\prime}(\rho)\left(\frac{u(\rho)+1}{\rho}-\rho\text{ }G\text{ }\left(\frac
{m^{2}\phi(\rho)^{2}}{2}+j\text{ }(\rho)\phi(\rho)\nonumber \right. \right. \\
\hfill  \left. \left. +\frac{j(\rho)^{2}}{2}+\frac{e(\rho)-p(\rho)}{2}\right)\right),\\
0  \, = \\ 
-\left(\phi(\rho)+j(\rho)\right)\text{ }j^{\prime}(\rho)+p^{\prime}(\rho)+\frac{v^{\prime}(\rho)}{2v}\left(p(\rho)+e(\rho)\right) \, .
\end{multline}

In order to reduce the number of parameters in the equations, let us define
a new radial variable, scalar field and other parameters as follows
\begin{align}
r  &  =m\rho,\\
\Phi(r)  &  =\sqrt{8\pi}l_{p}\phi(\rho) \, ,\\
J(r)  &  =\frac{\sqrt{8\pi}l_{p}}{m^{2}}j(\rho) \, ,\\
\text{\ \ }\epsilon(r)  &  \equiv\frac{8\pi l_{p}^{2}}{m^{2}}e(\rho) \, ,\\
P^{\ast}(r)  &  =\frac{8\pi l_{p}^{2}}{m^{2}}p(\rho) \, .
\end{align}

With the new notation the working equations  become
\begin{multline}
{\frac{u_{,r}(r)}{r}}-{\frac{1-u(r)}{r^{2}}} \,  = \\
-\frac{1}{2}\left(u(r){\Phi
_{,r}(r)}^{2}+\Phi(r)^{2}+2J(r)\Phi(r)+J(r)^{2}\right)-\epsilon(r),\label{e1}
\end{multline}
\begin{multline}
\frac{u(r)}{v(r)}\frac{v_{,r}(r)}{r}-{\frac{1-u(r)}{r^{2}}} \,  = \\
-\frac{1}{2}\left(-u(r){\Phi_{,r}(r)}^{2}+\Phi(r)^{2}+2J(r)\Phi(r)+J(r)^{2})-P^{\ast}(r)\right),\label{e2}
\end{multline}
\begin{multline}
0  \, = \\
P^{\ast\prime}(\rho)+\left(\epsilon(r)+P^{\ast}(r)\right)\frac{v_{,r}(r)}{2v(r)}-\left(J(r)+\Phi(r)\right)J_{,r}(r),\label{e3}
\end{multline}
\begin{multline}
J(r)+{\Phi}(r)-u(r)\text{ }{\Phi}^{\prime\prime}(r)  \, = \\ 
{\Phi}^{\prime}(r)\left(\frac{u(r)+1}{r}-r\text{ }\left(\frac{{\Phi}(r)^{2}}{2}+J(r){\Phi}(r) \right. \right.\label{e4}\\
  \hfill \left. \left. +\frac{J(r)^{2}}{2}+\frac{\epsilon(r)-P^{\ast}(r)}{2}\right)\right).%
\end{multline}

In order to simplify the notation, the same letters $u$ and $v$ have 
been used to indicate the metric components in the new variables. That is we
write $u(r)=u(\rho)$ and $v(r)=v(\rho)$ in spite of the fact that
functional forms of the two letters can not be equal. This should not create confusion.

\end{document}